\documentclass[11pt]{article}
\usepackage{moriond}
\usepackage{xspace}
\usepackage{amsmath}

\ifx\pdfoutput\undefined
\usepackage[dvips]{graphicx}
\else
\usepackage[pdftex]{graphicx}
\usepackage{epstopdf}
\fi

\setkeys{Gin}{width=.9\columnwidth, height=.37\textheight, keepaspectratio}

\def\Journal#1#2#3#4{{#1} {\bf #2}, #3 (#4)}

\def\PRD{Phys. Rev. D}

\def\be{\begin{equation}}
\def\ee{\end{equation}}
\def\bea{\begin{eqnarray}}
\def\eea{\end{eqnarray}}

\newenvironment{cfigure}[1][tbp]{\begin{figure}[#1]\centering}{\end{figure}}
\newcommand{\fig}[1]{Fig.~\ref{#1}}

\newcommand{\Fig}[1]{Figure~\ref{#1}}

\newcommand{\ppbar}  {\ensuremath{p\bar{p}}\xspace}
\newcommand{\ttbar}  {\ensuremath{t\bar{t}}\xspace}
\newcommand{\mreco}  {\ensuremath{m_{t}^{\text{reco}}}\xspace}
\newcommand{\mtop}   {\ensuremath{M_{\text{top}}}\xspace}
\newcommand{\mjj}    {\ensuremath{m_{\mathrm{jj}}}\xspace}
\newcommand{\jes}    {\ensuremath{\Delta_{\mathrm{JES}}}\xspace}
\newcommand{\jesmeat}    {\ensuremath{JES}\xspace}
\newcommand{\tev}[1]{\ensuremath{#1~\mathrm{TeV}}}
\newcommand{\gevcc}[1]{\ensuremath{#1~\mathrm{GeV}/c^{2}}}

\newcommand{\invpb}[1]{\ensuremath{#1~\mathrm{pb}^{-1}}}

\newcommand{\invfb}[1]{\ensuremath{#1~\mathrm{fb}^{-1}}}

\newcommand{\measErr}[2]{\ensuremath{#1 \pm #2}\xspace}
\newcommand{\sigcunit}[1]{\ensuremath{#1~\sigma_c}\xspace}
\newcommand{\pt}     {\ensuremath{p_{T}}\xspace}
\newcommand{\chisq}  {\ensuremath{\chi^{2}}\xspace}

\newcommand{\measStatSyst}[3]{\ensuremath{#1 \pm #2~(\text{stat.}) \pm #3~(\text{syst.})}\xspace}
\newcommand{\measAStatSyst}[4]{\ensuremath{#1~^{+#2}_{-#3}~(\text{stat.})\pm #4~(\text{syst.})}\xspace}

\begin{document}
\vspace*{4cm}
\title{TOP QUARK MASS MEASUREMENTS AT CDF}

\author{ERIK BRUBAKER, FOR THE CDF COLLABORATION}

\address{The Enrico Fermi Institute,\\
The University of Chicago,\\
Chicago, Illinois 60637, USA}

\maketitle\abstracts{
The mass of the top quark \mtop is interesting both as a fundamental parameter of
the standard model and as an important input to precision electroweak tests.
The Collider Detector at Fermilab (CDF) has a robust program of top quark mass
analyses, including the most precise single measurement,
$\mtop=\gevcc{\measErr{173.4}{2.8}}$, using \invpb{680} of \ppbar collision data.
A combination of current results from CDF gives
$\mtop=\gevcc{\measErr{172.0}{2.7}}$, surpassing the stated goal of \gevcc{3} precision
using \invfb{2} of data.  Finally, a combination with current D0 results gives
a world average top quark mass of \gevcc{\measErr{172.5}{2.3}}.}

\section{Introduction}
The top quark, as the heaviest known fundamental particle, is especially interesting
because of its large mass.  Top quark loops contribute large radiative
corrections to other observables such as the $W$ boson mass, and the value of
those corrections depends strongly on the top quark mass.  In particular, precise
measurements of the top quark and $W$ boson masses are needed to constrain the mass
of the putative Higgs boson, and for consistency studies if the Higgs is observed.
The Yukawa coupling of the top quark to a standard model Higgs is roughly one,
possibly indicating a special role for top in electroweak symmetry breaking.
Finally, the mass of the top quark has inherent interest as a fundamental parameter
of the standard model.

\section{Top Quark Mass at CDF}
The Collider Detector at Fermilab (CDF) is a general-purpose detector at the
Fermilab Tevatron, a \ppbar collider operating at $\sqrt{s}=\tev{1.96}$.
Top quark mass analyses at CDF use events where top/antitop pairs are produced
strongly.  Since top quarks decay before hadronizing to a $W$ boson and a
$b$ quark, the \ttbar event signature depends on the decays of the two $W$ bosons.
Here I report on measurements using the ``lepton + jets'' decay topology, where one
$W$ decays to an electron or muon and a neutrino, and the other $W$ decays
to hadrons; and the ``dilepton'' decay topology, where both $W$ bosons decay to $e\nu_e$
or $\mu\nu_{\mu}$.  Analyses using the challenging ``all-hadronic'' channel are in
progress at the time of this conference; no CDF top quark mass analyses select
$t\rightarrow Wb\rightarrow\tau\nu_{\tau}b$ at this time.  In some analyses, the sample
purity is significantly improved by requiring at least one $b$ tag, namely a
displaced secondary vertex inside a jet.

Extracting the top quark mass from these events is challenging for several reasons.
First, the events are complicated.  Due to the splitting and merging of jets, and the
presence of initial-state and final-state radiation (ISR \& FSR), only in
about 50\% of selected lepton + jets channel events do the four leading jets
correspond to the four quarks in the \ttbar decay chain.  Even if the correct
set of jets is selected, it is not easy to determine which jet corresponds to which
quark; the presence of $b$ tags eases this task.
In the case of dilepton events, the two neutrinos are observed as a single
combined ``missing energy'' vector in the plane transverse to the beam.  Second, jets
dominate \ttbar events as primary and secondary decay products of the top quarks.
But jet energies are measured with a relatively poor resolution of roughly
$85\%/\sqrt{E_{T}}$, contributing to statistical uncertainties in the top quark mass;
more importantly, there are large systematic uncertainties on the jet energy scale
that induce the dominant systematic uncertainty on the top quark mass. \cite{r:jesnim}
Third, the
selected event samples have background contamination at the few percent to fifty
percent level, depending on the cuts.  The background contributions are generally
well understood, but must be treated properly to minimize associated biases and
uncertainties.

CDF has a robust program of top quark mass measurements.  Techniques
with different sensitivity to event information and systematic effects yield
compatible results, giving us confidence in their accuracy.  About ten different
analysis techniques have been used at CDF on run II data; here we will concentrate on
recent results, all using 680--\invpb{750} of data.

The analysis techniques used by CDF to extract the top quark mass can be separated
into two categories.  Template methods choose some kinematic observable from each
event, often an event-by-event reconstructed top quark mass.
Monte Carlo samples with full detector simulation are used to build
``templates'' of the distribution of this variable for signal samples
generated with various values of the top quark mass, and for background processes.
A likelihood fit of the data distribution to probability density functions (p.d.f.'s)
derived from the signal and background templates yields a measurement of the
true top quark mass.  Due to this two-step process, the chosen observable does not
need to be an unbiased estimator of the true mass, and in fact the template shapes
are often quite complicated (see \fig{f:tmtdistos} for an example).
Matrix element methods, in contrast, use a direct expression of the probability that
a particular event is observed given a true top quark mass.  The likelihood is built
up from parton distribution functions, matrix elements for signal and background
processes, and ``transfer functions'' that connect quarks at the matrix element level
to observed jets, accounting for fragmentation effects and detector resolution.
The measured top quark mass is then the maximum of the product of all the event
likelihood curves. Since various approximations must be made in the interest of
computational tractability, the measurement must in general be calibrated using
the observed behavior on fully simuluated Monte Carlo samples with known top quark
mass.  Thus both sorts of analyses have as an essential input a good Monte Carlo
description of the physics processes and a good detector simulation.

\section{Top Quark Mass Analyses With 680--750 pb$^{-1}$}

In the current data sample of \invpb{750}, our standard selection cuts \cite{r:selcuts}
select 64 dilepton and 360 lepton + jets \ttbar candidates.  In the \invpb{680}
subset of the data where the CDF silicon microstrip detector was operating well,
allowing secondary vertices to be reconstructed, we find 27 dilepton and
252 lepton + jets candidate events with at least one $b$ tag.  These are the
largest \ttbar datasets yet analyzed, and recent measurements of the top quark
mass have statistical uncertainties significantly improved over those shown
at previous conferences.

\subsection{Two-dimensional Template Analysis in the Lepton + Jets Channel}
\label{ssec:tmt}

The jet energy scale systematic uncertainty is the dominant contribution
to systematic uncertainties on the top quark mass.  It is difficult to
calibrate jet energies at a hadron collider in general because resonant
decays to jets are impossible to separate from the continuum background,
leaving no point of reference for jet energy measurements.  But
in our reasonably high-purity sample of \ttbar events in the lepton + jets
channel, there exists a resonant decay of a $W$ boson to two jets.  CDF has
developed a technique to use this information to constrain our jet energy
scale uncertainty, which results in a lower systematic uncertainty on the
top quark mass. \cite{r:tmt2d}

This analysis is an extension of the traditional template method.  We would
like to measure or constrain two quantities, the true top quark mass \mtop
and the difference between our nominal jet energy calibration and the true
scale.  This difference, denoted \jes, is expressed in units of the
uncertainty in our nominal calibration, $\sigma_c$, so that we have by
definition a prior constraint of $\jes=\sigcunit{\measErr{0}{1}}$.  Instead
of a single reconstructed variable, we use two observables, the reconstructed
top quark mass \mreco and the dijet mass \mjj.  The former comes from a
kinematic fit to the full event, and the latter is the invariant mass of the
presumed $W$ daughter jets.

The quantity \mreco is primarily sensitive to the
value of \mtop, but is also affected by \jes.  The $W$ boson mass is known so
precisely that it can be treated as fixed, and the distribution
of \mjj is primarily sensitive to \jes.  But \mjj also has a dependence
on \mtop, since sometimes we erroneously assign one of the $b$ jets from top
decay as a $W$ daughter jet.  Given all the dependencies,
we fit simultaneously for the two quantities of interest, using the distributions
of both observables.  In addition, we subdivide the sample into
four categories of events, determined by the
number of $b$ tags and the jet transverse energies.  Since the subsamples have
different signal purity and different resolutions in the two observables,
treating them separately increases the power of the likelihood fit.

\begin{cfigure}
\includegraphics[width=.55\columnwidth]{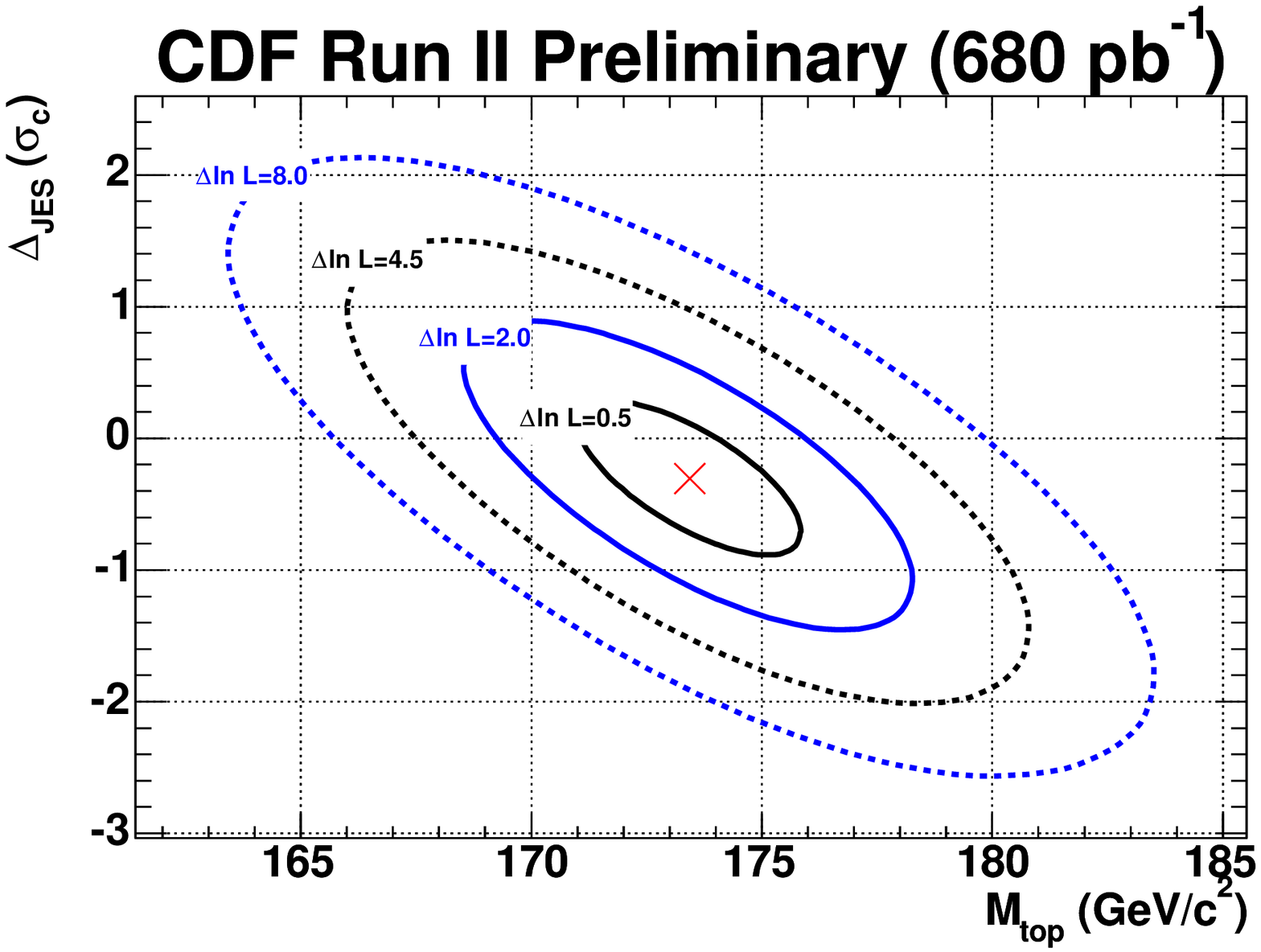}
\caption{
Results from the two-dimensional template method in the lepton + jets channel.
Contours are shown at various intervals of log-likelihood in the
\mtop-\jes plane; the maximum likelihood point is given by the red cross. 
}
\label{f:tmtcont}
\end{cfigure}

A global fit to the eight distributions (four subsamples times two observables),
using the above prior constraint on \jes, yields likelihood contours in the
\mtop-\jes plane as shown in \fig{f:tmtcont}.  The anticorrelation arises
because a given event sample can be consistent with a high top quark mass
and a low jet energy scale (i.e.\ jet energies reconstructed too low) or vice
versa.  The distributions of \mreco and \mjj, along with the predicted distributions
at the fitted values of \mtop and \jes, are shown in \fig{f:tmtdistos}.

\begin{cfigure}
\includegraphics[width=.45\columnwidth]{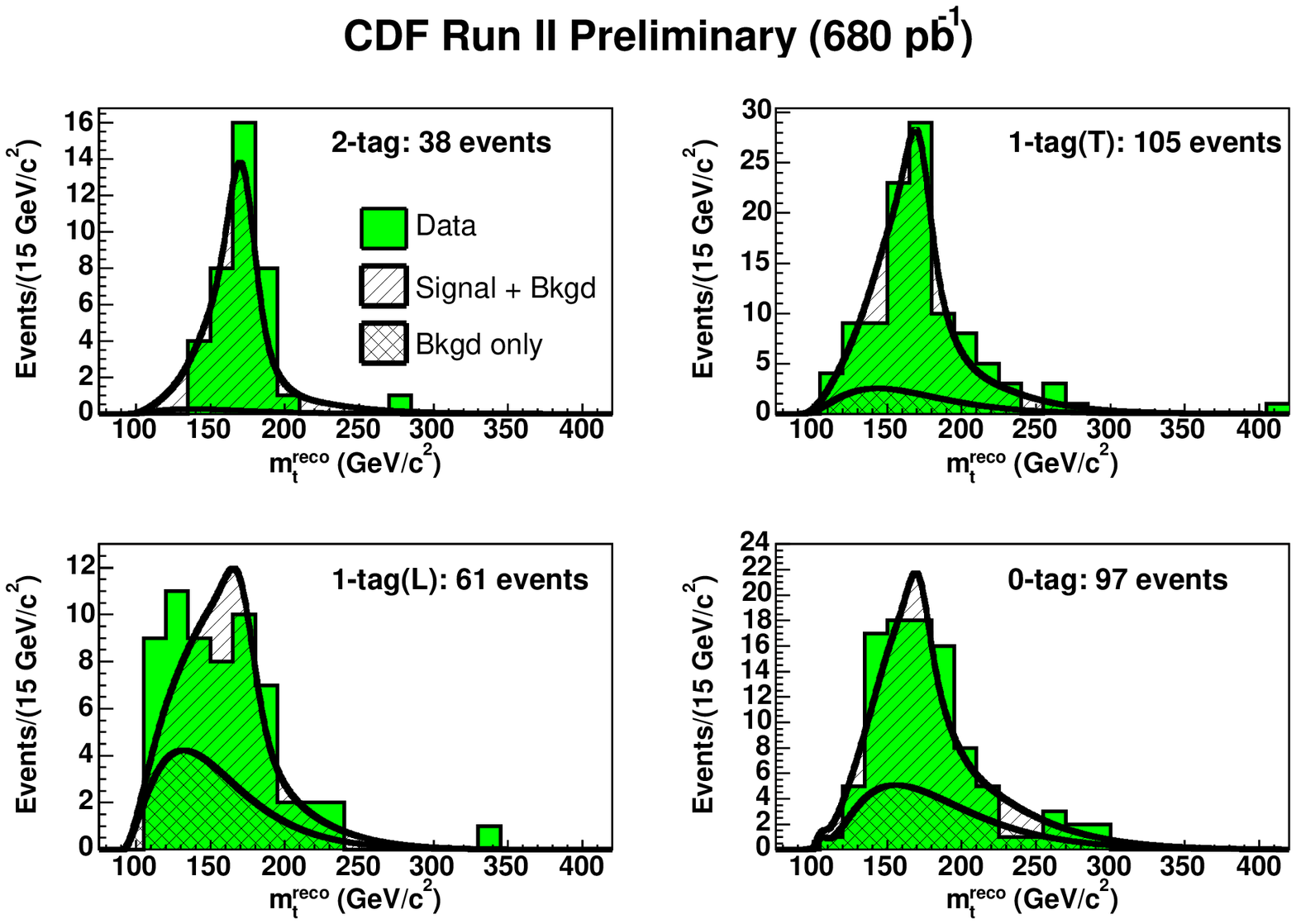}
\includegraphics[width=.45\columnwidth]{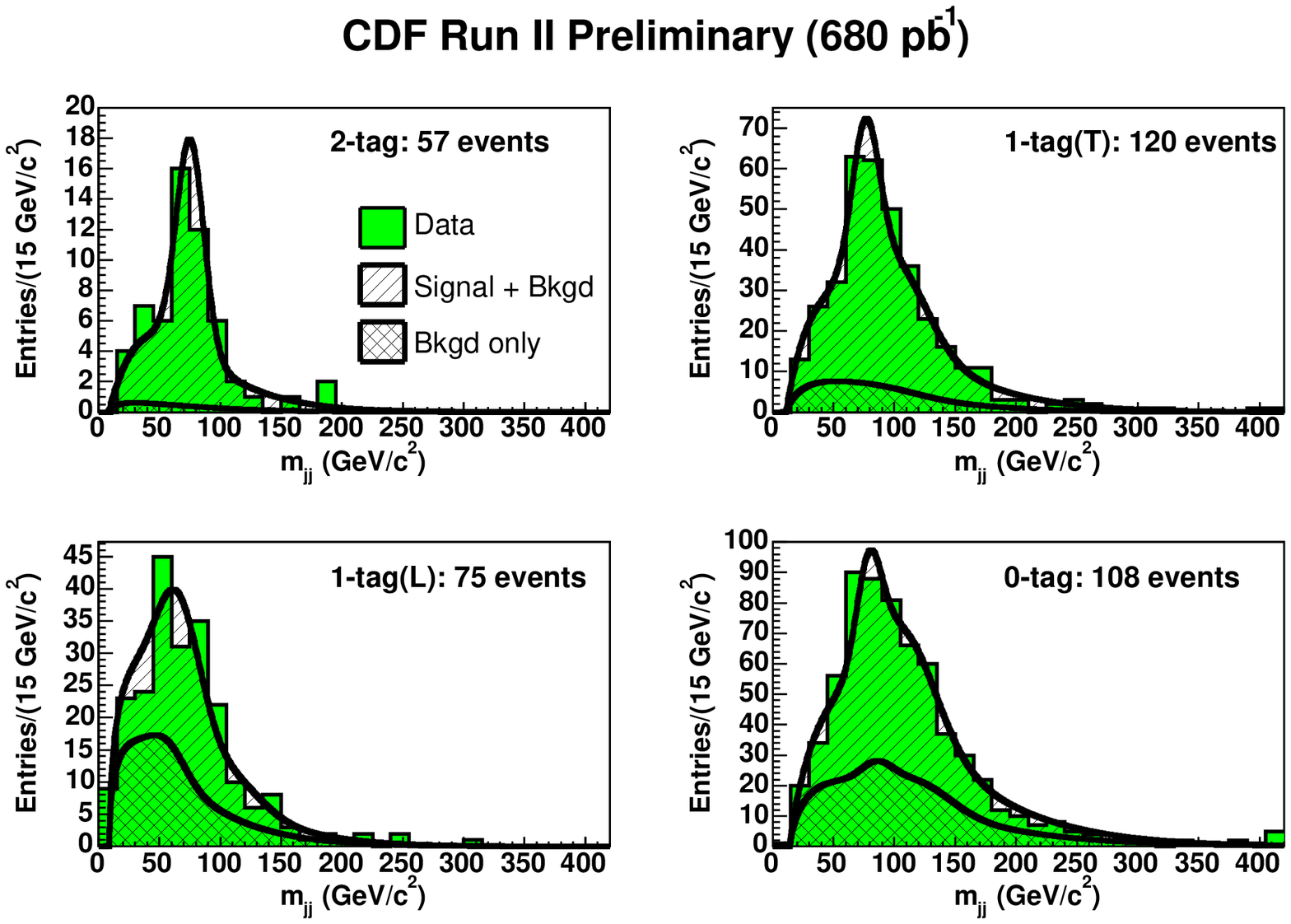}
\caption{
Results from the two-dimensional template method in the lepton + jets channel.
The plots show the distributions of \mreco (left), the reconstructed
top quark mass, and \mjj (right), the reconstructed dijet mass, for each of
the four categories of events.
In each case, the upper left plot is the subsample of events with
two $b$ tags, upper right is events with one $b$ tag and four high-\pt jets,
lower left is events with one $b$ tag and a relaxed cut on the fourth jet,
and the lower right is events with zero $b$ tags, but tighter jet \pt cuts.
}
\label{f:tmtdistos}
\end{cfigure}

The fitted value of \jes is $\sigcunit{\measErr{-0.3}{0.6}}$, which is consistent
with our prior understanding of jet energies, but directly translates into
a 40\% reduction in the corresponding systematic uncertainty on the top quark mass.
In fact, we have converted the dominant systematic error on \mtop into a
statistical error, which will improve as we accumulate more integrated
luminosity.  However, the remaining systematic uncertainties still get a large
contribution from components of the jet energy scale uncertainty that cannot
be accounted for with a single jet energy scale parameter.  Examples are the
relative uncertainties in our jet energy calibration between $W$ daughter jets
and $b$ jets, and between jets of different \pt.
Using the two-dimensional template method, the measured top quark mass
is $\mtop=\gevcc{\measStatSyst{173.4}{2.5}{1.3}}$, the most precise single measurement
of this important parameter.

\subsection{Two-dimensional Matrix Element Analysis in the Lepton + Jets Channel}
\label{ssec:meat}

We have recently completed a second analysis that simultaneously
measures the top quark mass and the jet energy scale, this time based on
a matrix element technique.  The likelihood to observe each event, generically
represented as a vector of observables $\vec{x}$, depends on the
true top quark mass, the jet energy scale \jesmeat (here defined as a simple
multiplicative factor on the jet energies), and the signal fraction $c_s$:
\begin{equation}
P_{0}(\vec{x};\mtop,\jesmeat,c_s)=c_{s}P_{\ttbar}(\vec{x};\mtop,\jesmeat)+
(1-c_{s})P_{\text{W+jet}}(\vec{x};\jes).
\end{equation}
The meat of the method is in the signal probability:
\begin{equation}
P_{\ttbar}(\vec{x};\mtop,\jesmeat)=\frac{1}{\sigma}\int d\sigma_{\ttbar}(\vec{y};
\mtop)dq_{1} dq_{2} f(q_{1}) f(q_{2}) W(\vec{x},\vec{y},\jesmeat).
\end{equation}
Here the differential cross-section $\sigma_{\ttbar}$, which is determined by
the \ttbar production and decay matrix element and depends on
parton-level kinematic quantities $\vec{y}$, is connected with the
observables $\vec{x}$ by the transfer function $W$, which incoporates the
dependence on the jet energy scale.  The variables $q_{1}$ and $q_{2}$ correspond
to the incoming partons, and $f$ represents the parton distribution functions.
The final measurement is based on the product of all the event likelihoods;
in this analysis the extrinsic jet calibration is not used, so all of the
jet energy scale information comes from the $W$ mass resonance.

In this method, the full kinematic and dynamical information from each event is
used, in contrast to the template method, where each event is reduced to two
numbers.  However, since a leading-order matrix element is used, events with
extra radiation are not well described, and the sample is restricted to events
with exactly four reconstructed jets.  Additionally, a requirement of at least
one $b$ tag is imposed to improve sample purity, leaving 118 events.

Since several approximations are made in the likelihoods as written above, we
calibrate the method against Monte Carlo samples as shown in the left plot
of \fig{f:meat}.  Within the statistical uncertainty of the Monte Carlo samples
used, the method is unbiased.  The right plot of \fig{f:meat} shows the likelihood
contours, with a result very similar to the template analysis above.  Although
the matrix element analysis uses fewer events, the more powerful method yields
a result of comparable sensitivity.  The systematic uncertainties are dominated
by uncertainties in the modeling of signal events, for example in the level of
ISR and FSR in \ttbar events.
Using the two-dimensional matrix element method, the measured top quark mass
is $\mtop=\gevcc{\measStatSyst{174.1}{2.5}{1.4}}$.

\begin{cfigure}
\includegraphics[width=.45\columnwidth]{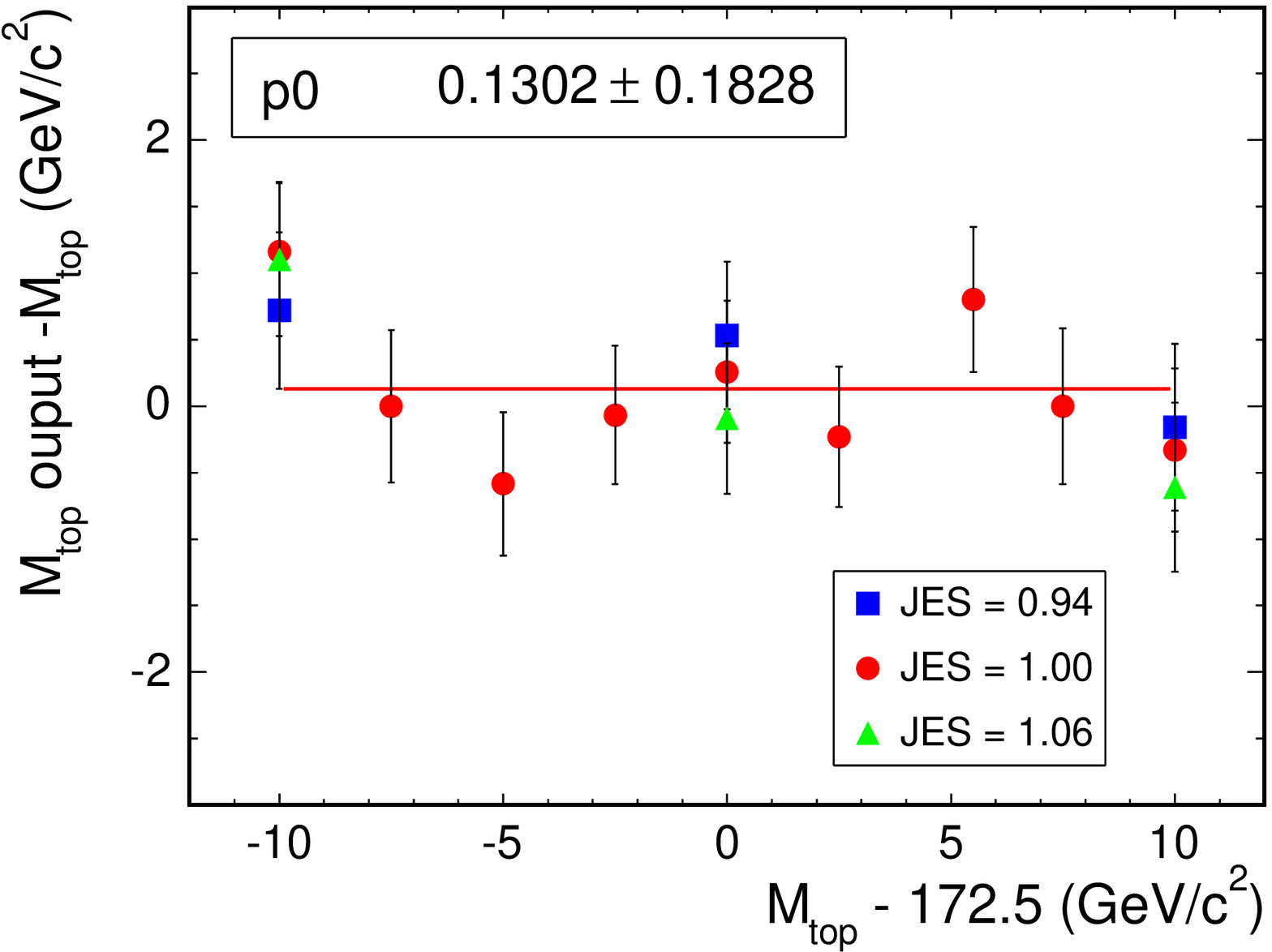}
\includegraphics[width=.48\columnwidth]{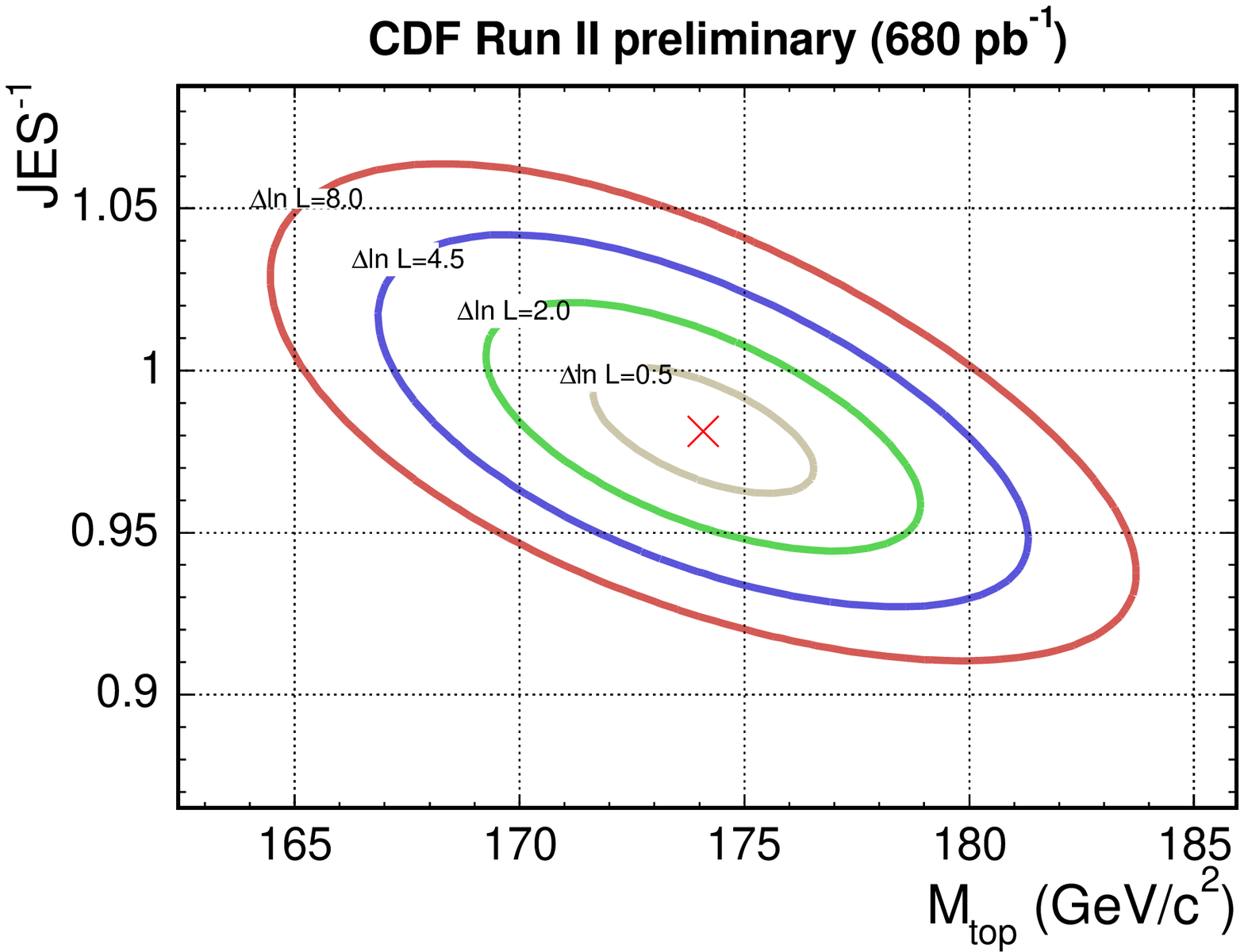}
\caption{
Calibration and result from the two-dimensional matrix element method
in the lepton + jets channel.  The left plot shows the difference between
the measured and generated \mtop in large Monte Carlo
samples with different generated
\mtop ($x$ axis) and different imposed values of \jes (legend).  The right
plot shows contours at various intervals of log-likelihood in the \mtop-\jesmeat
plane; the maximum likelihood point is given by the red cross.
}
\label{f:meat}
\end{cfigure}

\subsection{B Hadron Decay Length Analysis in the Lepton + Jets Channel}

In this template-based analysis we take a different approach to reducing
the jet energy scale systematic, namely we eliminate it by
disregarding the jet energies altogether.  Recall that the $b$ tagging
algorithm reconstructs a secondary vertex inside a jet, identifying the
displacement between the primary and secondary vertices with the flight
path of a $B$ hadron.  In \ttbar events, the precise magnitude of that
displacement depends on the boost imparted to the associated $b$ quark,
which in turn depends on the mass of its predecessor, the decaying top
quark.  Thus we choose as our observable the decay length in $b$-tagged
jets; the uncertainties on this quantity are uncorrelated with the jet
energy scale uncertainty.

In practice, this is a challenging measurement that is still limited by
the statistics of the event sample.  A proof-of-principle analysis has
been performed using lepton + jets events, but dilepton and even all-hadronic
events could be used in the future.  \Fig{f:lxy} shows the distribution
of the transverse decay length in the data, compared with the expected
contributions from signal and background events.  The fitted top quark mass
is $\mtop=\gevcc{\measAStatSyst{183.9}{15.7}{13.9}{5.6}}$, consistent with other
measurements within the large uncertainties.  An important point is that
the statistical and systematic uncertainties are almost completely
uncorrelated with other measurements, so that the method will soon begin
to contribute noticeably to \mtop combinations.  Studies show that
at the LHC, similar methods can make a significant contribution to the
overall top quark mass uncertainty. \cite{r:lxy}

\begin{cfigure}
\includegraphics[width=.55\columnwidth]{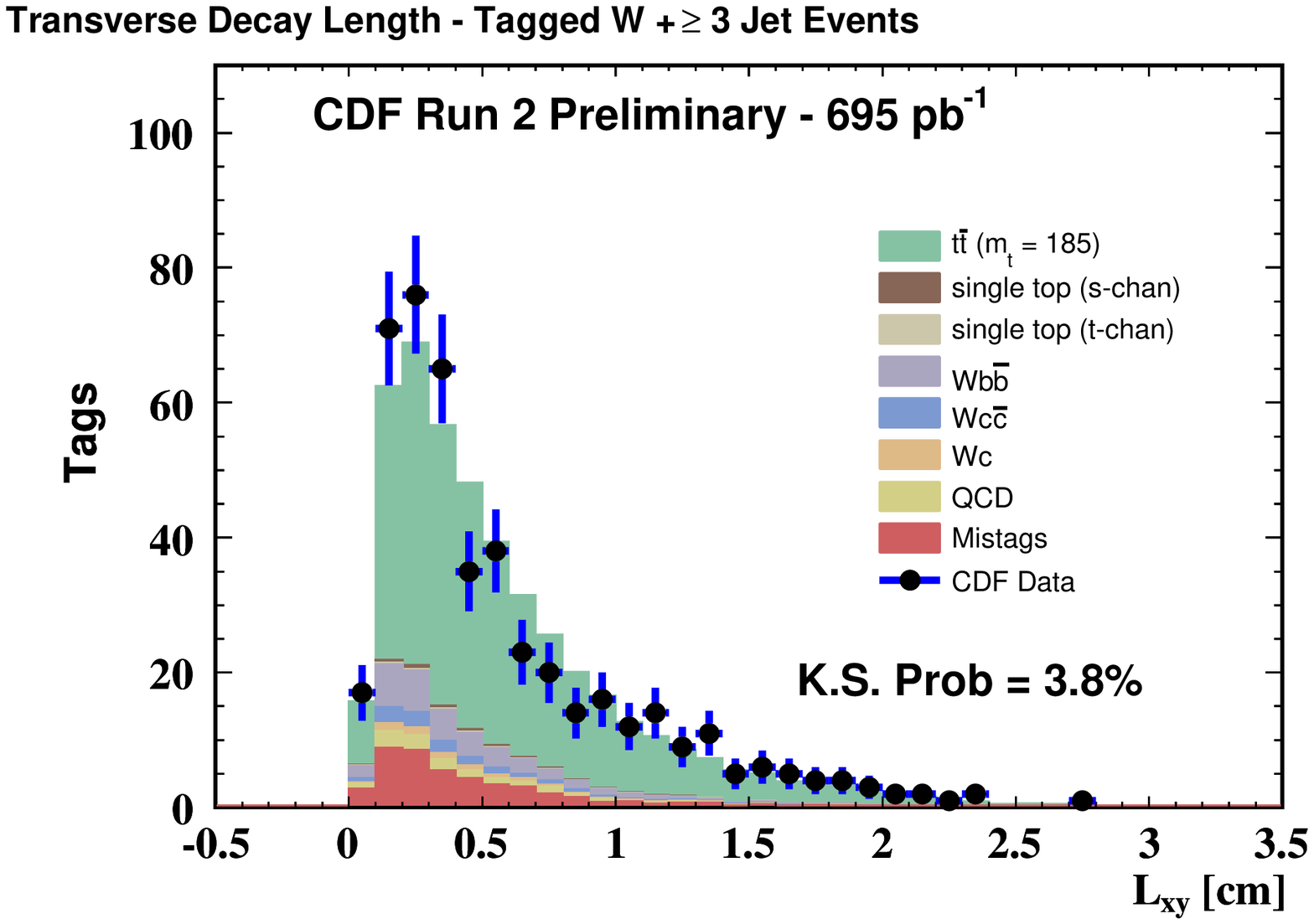}
\caption{
The transverse decay length $L_{xy}$, measured in $b$-tagged jets in
the lepton + jets channel,
is shown for the data compared with the expected contributions
from signal and various background processes. 
}
\label{f:lxy}
\end{cfigure}

\subsection{Matrix Element Analysis in the Dilepton Channel}
\label{ssec:madcow}

A matrix element technique is also applied to events in the dilepton
channel.  In dilepton events, the top quark mass is harder to reconstruct:
the two neutrinos result in an underconstrained kinematic system.  The
dilepton sample is thus particularly amenable to a matrix element approach,
since the observed quantities and their correlations are used in a more
optimal fashion.

We use a likelihood similar in form to that given above in Section~\ref{ssec:meat}.
In this case, however, the approximations have a more significant effect.  The left
plot of \fig{f:madcow} shows the calibration curve derived from Monte Carlo samples
generated with various values of \mtop.  The slope of this curve is \measErr{0.85}{0.01},
indicating a substantial nonlinearity in the raw measurement, which
comes primarily from limitations in the background modeling.
Additionally, the pull widths for the raw measurement are about 1.5 (with no dependence
on true top mass), indicating a significantly underestimated uncertainty.  This effect
has been studied extensively, and is understood to result from the assumptions that
i) lepton energies are perfectly measured; ii) lepton and jet angles are perfectly
measured; iii) the leading two jets in a signal event always correspond to the two $b$ jets
from top decay; and iv) the background events are well modeled.  After calibration, the
measurement is unbiased and the error is accurately estimated over a large range of \mtop.

\begin{cfigure}
\includegraphics[width=.5\columnwidth]{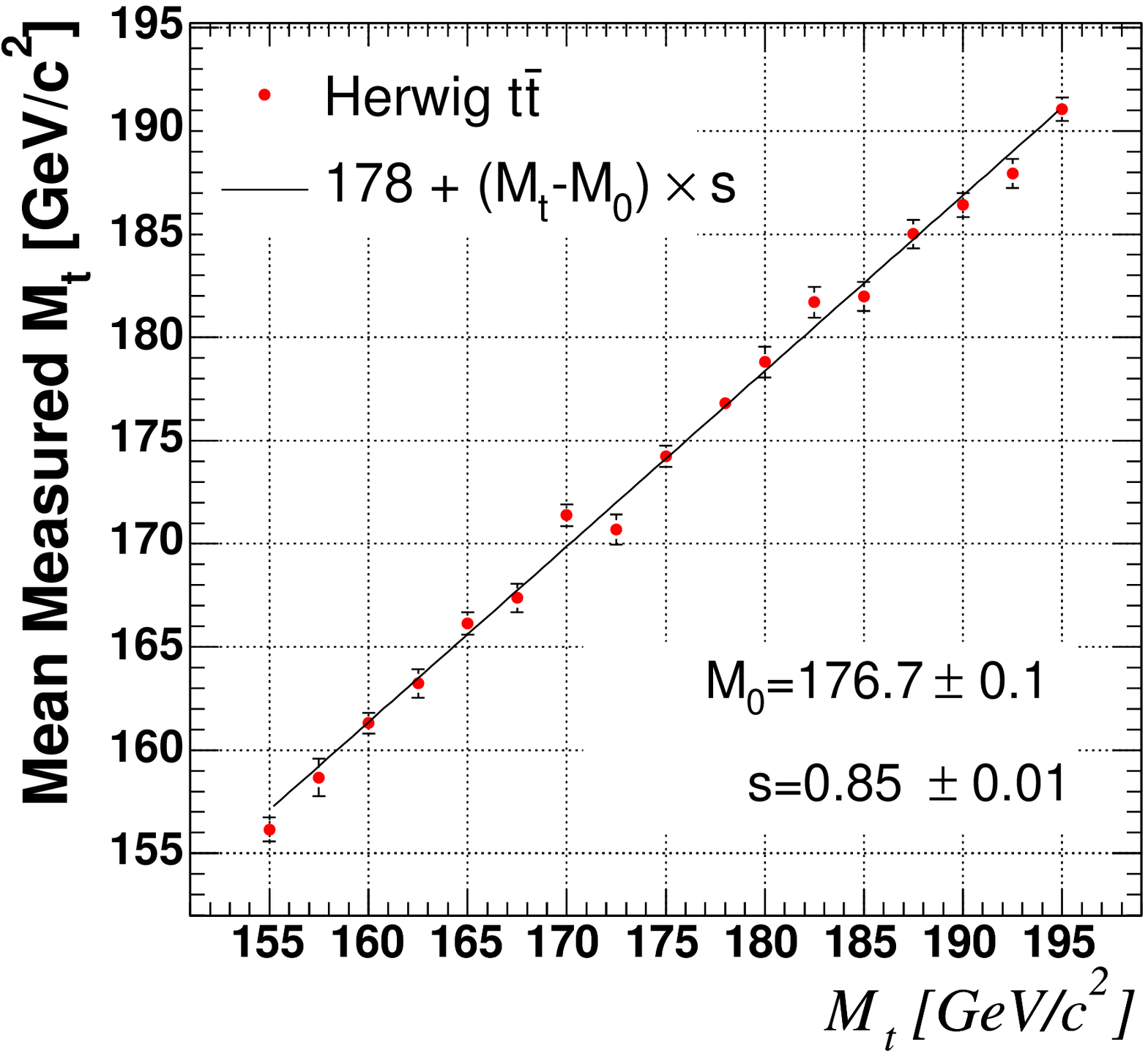}
\includegraphics[width=.45\columnwidth]{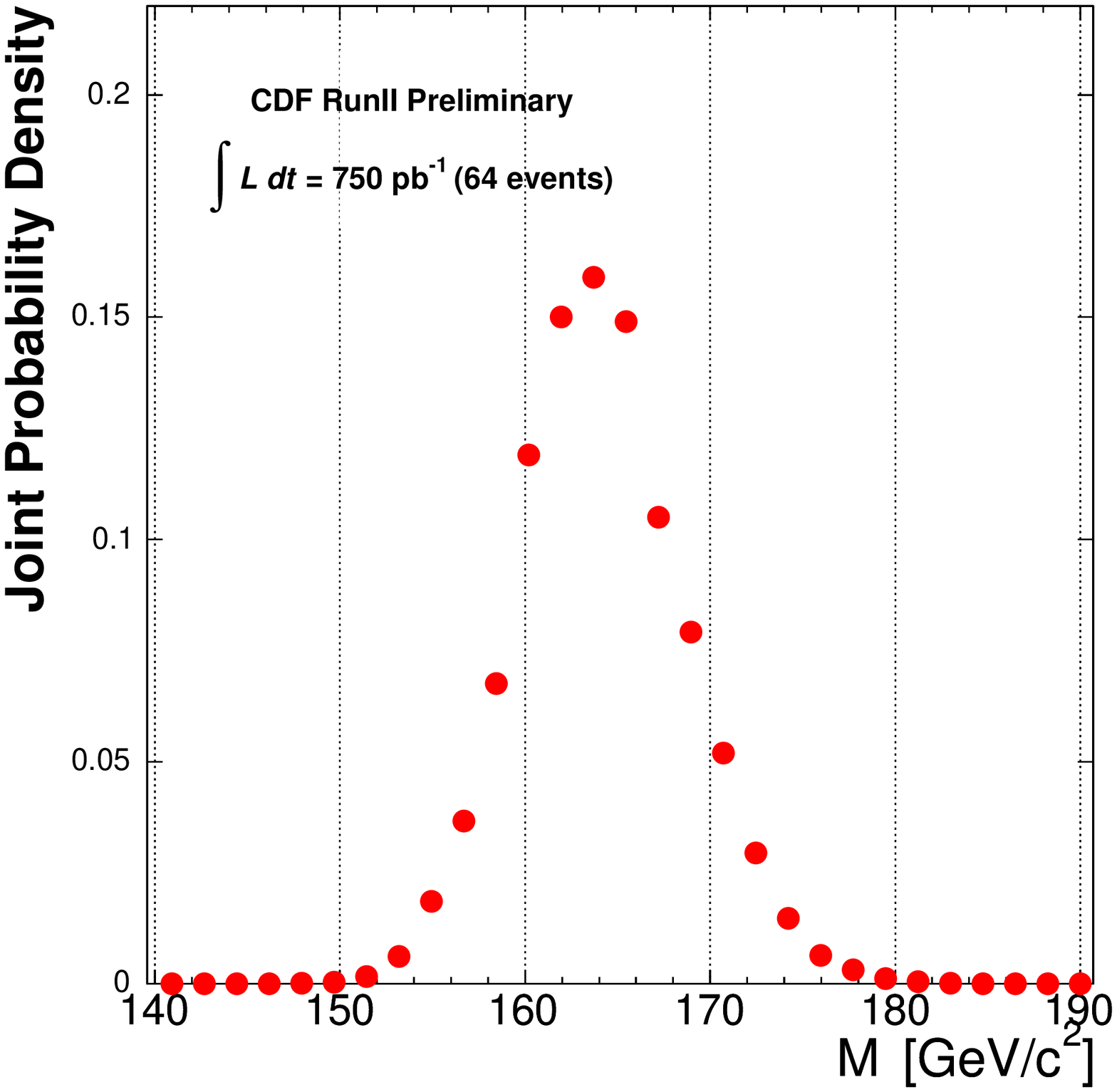}
\caption{
Calibration and result from the matrix element method
in the dilepton channel.  The left plot shows the calibration curve of
measured vs generated \mtop in large Monte Carlo samples.  The right
plot shows the joint likelihood curve for the 64 selected data events.
}
\label{f:madcow}
\end{cfigure}

The right plot of \fig{f:madcow} shows the resulting likelihood for the 64 events in the
dilepton sample.  From these events we measure $\mtop=\gevcc{\measStatSyst{164.5}{4.5}{3.1}}$,
the most precise single measurement in this challenging channel.  The measurement is
also performed on the subset of 27 events with at least one $b$ tag, yielding
$\mtop=\gevcc{\measStatSyst{162.7}{4.6}{3.0}}$.  In both cases, the systematic uncertainty is
dominated by the jet energy scale, which contributes about \gevcc{2.5}.

\section{Combining Top Quark Mass Measurements}

The most precise information about the top quark mass comes from combining the
measurements from different analyses, taking advantage of the fact that each method
is sensitive to different parts of the total available experimental information.  We
use the BLUE (best linear unbiased estimator) method \cite{r:blue} to correctly account for
the correlations among the systematic uncertainties in different analyses.  To date,
we combine only one analysis from each channel, although work is ongoing to understand
the statistical correlation between analyses using the same events so that more
measurements can be included in the combination.

\subsection{CDF Combination}

For this conference, we performed a combination of the CDF run I measurements and the
measurements described in Sections~\ref{ssec:tmt} and~\ref{ssec:madcow} of this
document.  The combined top quark mass is
$\mtop=\gevcc{\measStatSyst{172.0}{1.6}{2.2}}=\gevcc{\measErr{172.0}{2.7}}$,
exceeding the precision
of the previous world average.  The inputs are consistent when all correlations are
taken into account: $\chisq/ndof=5.1/4$, with a p-value of 28\%.  CDF's goal for
top quark mass precision in run IIa (\invfb{2}) was \gevcc{3}.

\Fig{f:cdfextrap_mtmw}, left side, shows an extrapolation of CDF's past performance
on top quark mass measurements into the future.  Two scenarios are given, one where
systematic uncertainties are fixed and only statistical uncertainties improve with
more data; the second where systematic uncertainties continue to improve at the same
rate as statistical uncertainties, as has been the case in the past.  The real
future of CDF top quark mass measurements probably lies somewhere between the two lines.

\begin{cfigure}
\includegraphics[width=.45\columnwidth]{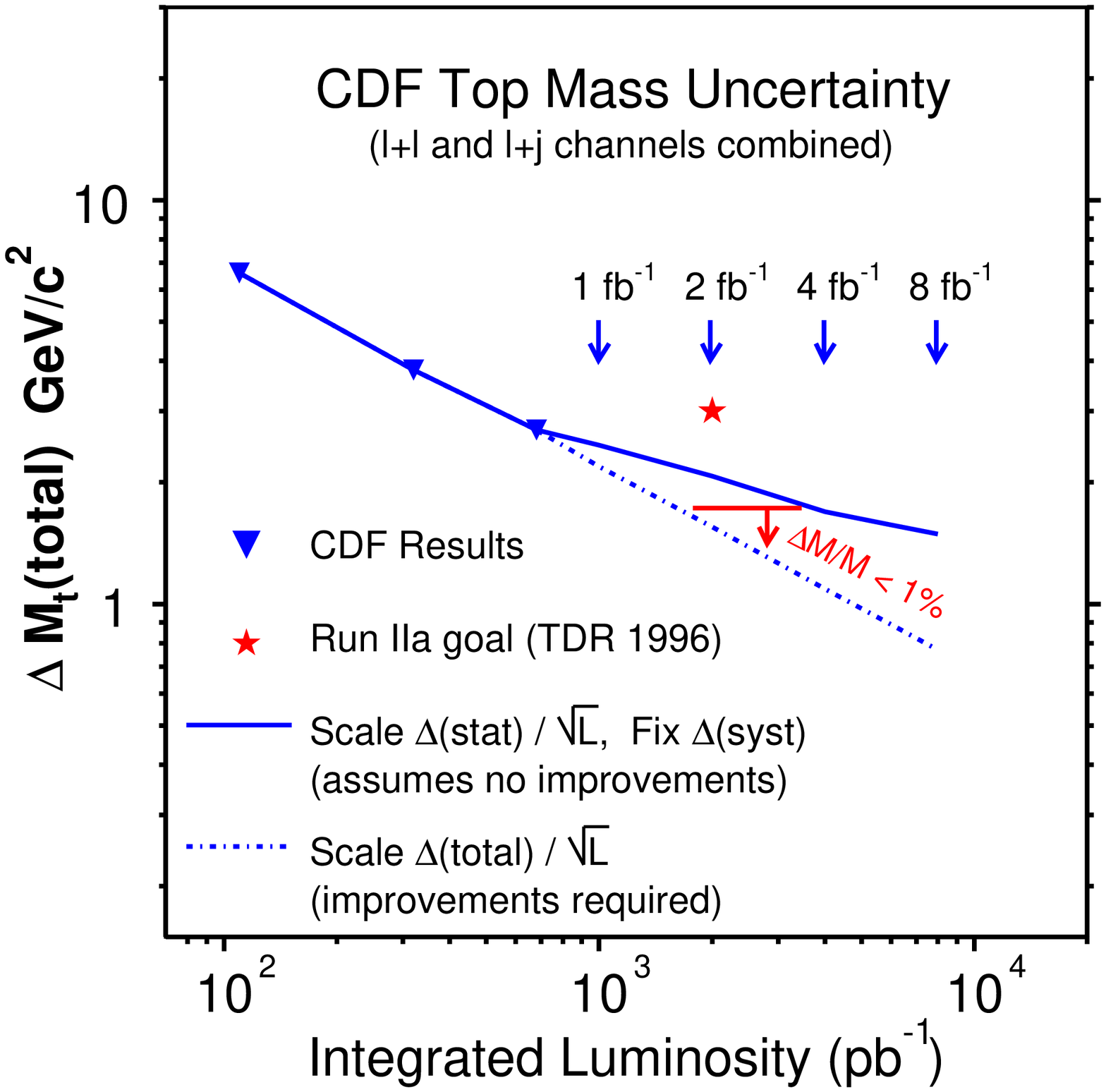}
\includegraphics[width=.45\columnwidth]{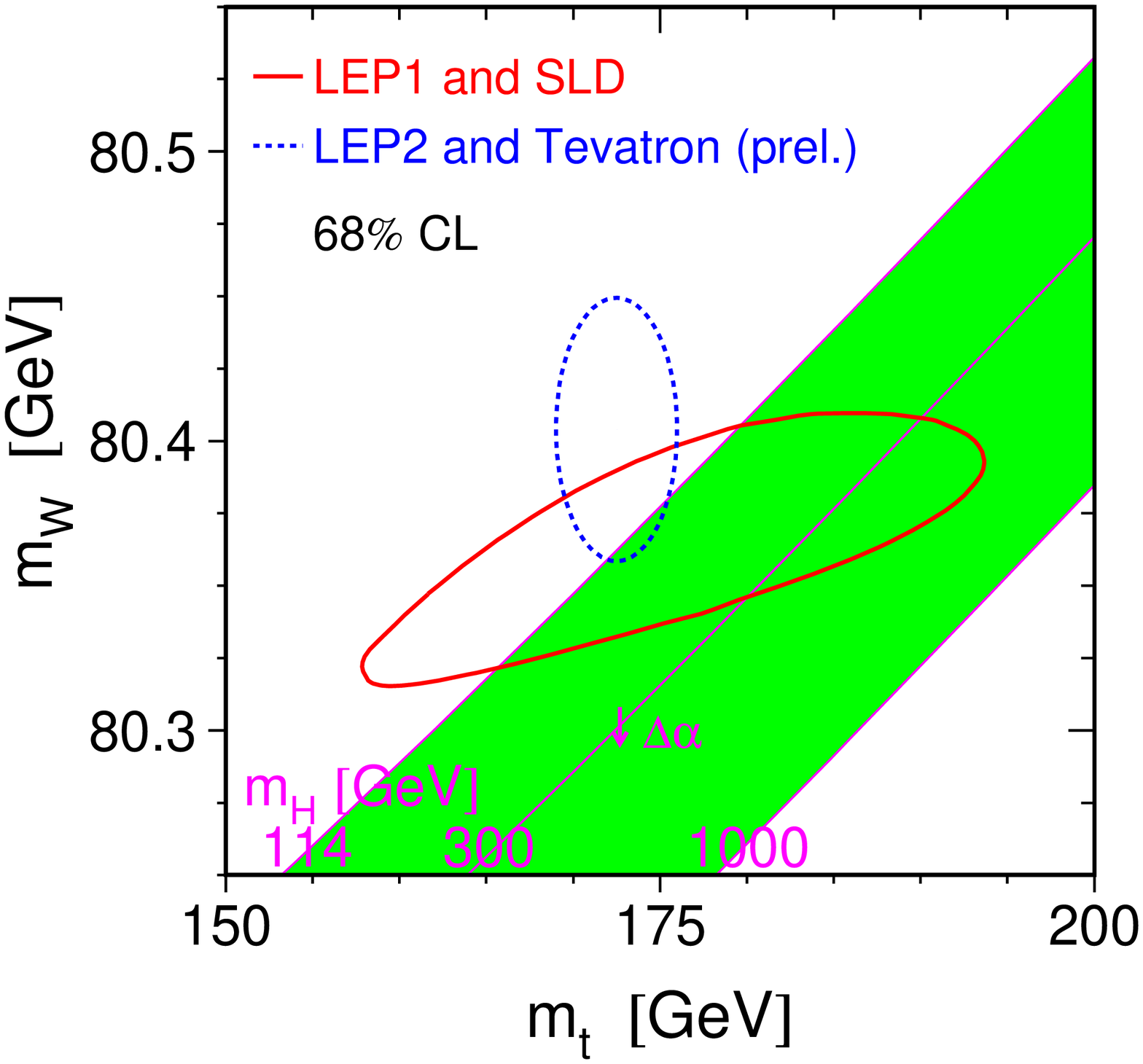}
\caption{In the left plot, the precision of past CDF top quark mass
measurements is plotted as a function of the integrated luminosity
on a log-log scale.
An extrapolation to the future assuming no improvement in systematic
uncertainties is shown with a solid line; if systematics
continue to be reduced as in the past, the dashed line would result.
The right plot summarizes the current status of the global electroweak
fits in the context of the standard model, using the new world average
top quark mass value produced for this conference.
Direct and indirect measurements of the top quark and $W$ boson masses
are shown using 68\% CL contours, and bands indicate the preferred Higgs
boson mass at each point in the \mtop-$M_{W}$ plane.
}
\label{f:cdfextrap_mtmw}
\end{cfigure}

\subsection{Tevatron Combination}

A combination of current CDF and D0 measurements was performed for this
conference using the same method. \cite{r:tevcomb}
The new world average top quark mass using
preliminary measurements from both collaborations is
$\mtop=\gevcc{\measStatSyst{172.5}{1.3}{1.9}}=\gevcc{\measErr{172.5}{2.3}}$.
The effect on the standard model electroweak precision fit is shown in
the right plot of \fig{f:cdfextrap_mtmw}, which uses the new combination.

\section{Conclusion}

CDF has a strong program of top quark mass measurements, which has recently
been augmented by a new set of results using \invpb{680-750}.  A new combination
of current CDF results yields a top quark mass measurement of
\gevcc{\measErr{172.0}{2.7}}, putting 1\% precision on this important parameter
in reach as the experiment continues to accumulate integrated luminosity.

\section*{Acknowledgments}
We thank the Fermilab staff and the technical staffs of the participating
institutions for their vital contributions. This work was supported by the
U.S. Department of Energy and National Science Foundation; the Italian
Istituto Nazionale di Fisica Nucleare; the Ministry of Education, Culture,
Sports, Science and Technology of Japan; the Natural Sciences and
Engineering Research Council of Canada; the National Science Council of
the Republic of China; the Swiss National Science Foundation; the A.P.
Sloan Foundation; the Bundesministerium f\"ur Bildung und Forschung,
Germany; the Korean Science and Engineering Foundation and the Korean
Research Foundation; the Particle Physics and Astronomy Research Council
and the Royal Society, UK; the Russian Foundation for Basic Research;
the Comisi\'on Interministerial de Ciencia y Tecnolog\'{\i}a, Spain;
in part by the European Community's Human Potential Programme under
contract HPRN-CT-2002-00292; and the Academy of Finland.

\end{document}